\documentclass[aps,twocolumn,showpacs,superscriptaddress,amsmath,amssymb,floatfix,longbibliography]{revtex4-1}
\usepackage[breaklinks=true,colorlinks,citecolor=blue,linkcolor=blue,urlcolor=blue]{hyperref}
\usepackage{float}
\usepackage{physics}
\usepackage{epsfig,mathrsfs,color,latexsym,subfigure,marginnote}
\usepackage{graphicx,longtable,amsmath,amssymb,amsfonts}
\usepackage{multirow}
\usepackage{graphicx}

\usepackage{ulem}

\begin{document}

\title{Topological nonsymmorphic insulator versus Dirac semimetal in KZnBi}

\author{Rahul Verma}
\affiliation{Department of Condensed Matter Physics and Materials Science, Tata Institute of Fundamental Research, Mumbai 400005, India}

\author{Bikash Patra}
\affiliation{Department of Condensed Matter Physics and Materials Science, Tata Institute of Fundamental Research, Mumbai 400005, India}

\author{Bahadur Singh}
\email{bahadur.singh@tifr.res.in}
\affiliation{Department of Condensed Matter Physics and Materials Science, Tata Institute of Fundamental Research, Mumbai 400005, India}

\begin{abstract}
KZnBi was discovered recently as a new three-dimensional Dirac semimetal with a pair of bulk Dirac fermions in contrast to the $\mathbb{Z}_2$ trivial insulator reported earlier. In order to address this discrepancy, we have performed electronic structure and topological state analysis of KZnBi using the local, semilocal, and hybrid exchange-correlation (XC) functionals within the density functional theory framework. We find that various XC functionals, including the SCAN meta-GGA and hybrid functional with 25\% Hartree-Fock (HF) exchange (HSE06), resolve a topological nonsymmorphic insulator state with the glide-mirror protected hourglass surface Dirac fermions. By carefully tuning the XC strength in modified Becke-Johnson (mBJ) potential, we recover the correct orbital ordering and Dirac semimetal state of KZnBi. We further show that increasing the default HF exchange in hybrid functional ($> 40\%$) can also capture the desired Dirac semimetal state with the correct orbital ordering of KZnBi. The calculated energy dispersion and carrier velocities of Dirac states are found to be in excellent agreement with the available experimental results. Our results demonstrate that KZnBi is a unique topological material where large XC effects are crucial to producing the Dirac semimetal state.
\end{abstract}

\maketitle
\clearpage
\section{Introduction}

Interest in symmetry-protected nontrivial states has driven search and discovery efforts for finding materials with nontrivial electronic properties useful for fundamental studies and device applications~\cite{Singh2022,RMP_Kane2010,RMP_Ashvin}. Many topological materials are now predicted in high-throughput materials searches ranging from insulators to metals and semimetals with nontrivial states protected not only by the free-fermion symmetries such as time-reversal or particle-hole symmetries but also by crystalline symmetries~\cite{Singh2022,RMP_Kane2010,RMP_Ashvin,TDB_Tang2019,TDB_Vergniory2019,TDB_Zhang2019,TCI_Liangfu,TCI_LiangfuSnTe,TCI_superconductor,TCI_BSTaAs2}. Among various topological materials, three-dimensional (3D) Dirac semimetals support four-fold band crossings described by pseudo-relativistic Dirac fermions with linear energy dispersion in their electronic structure. These band crossings are protected by crystalline symmetries such as rotational or nonsymmorphic symmetries in the presence of time-reversal and inversion symmetries~\cite{DSM_3D_PRL,DSM_A3Bi,Yang2014,DSM_Na3Bi_SuYang,DSM_Cd3As2,DSM_CaAuAs}. Systematically reducing these symmetries can transition a topological Dirac semimetal into a nontrivial insulator or Weyl semimetal~\cite{Phase_Murakami,DSM_A3Bi,Weyl_Shinming,Weyl_ThBiSe2,DSM_BaAgAs}. The nonsymmorphic crystalline symmetries that accompany a symmetry operation with a fractional translation can protect topological states in nontrivial insulators. Such topological nonsymmorphic insulators can realize special hourglass-like Dirac fermions over the material's surface that respect nonsymmorphic symmetries~\cite{TNCI,Alex_groupCho,Hourglass_bernevig,KHgSb_Ma_exp,KHgSb_Liang_exp,HG_Ag2BiO3,HG_stacked_TI}. 

Among various families of topological materials, the AB$X$-type hexagonal honeycomb materials attract significant attention due to their greater materials and topological state tunability~\cite{PRL_2011,yan2012prediction,zunger_TIvsTDS, ABC_Polar, ABC_ferroelectric}. Specifically, these materials can realize $\mathbb{Z}_2$ topological insulators, topological nonsymmorphic crystalline insulators, topological Dirac semimetals, and three-dimensional quantum spin-Hall insulators as well as provide an opportunity to realize topological phase transitions from a nontrivial insulator state to a Dirac or Weyl semimetal state by pressure or systematically lowering the crystalline symmetries. For example, KHg$X$ ($X$ = As, Sb, or Bi) materials with Hg$X$ honeycomb layer host topological non-symmorphic crystalline insulator with hourglass-like surface Dirac fermions and undergo a phase transition to topological Dirac semimetal state under external pressure~\cite{Hourglass_bernevig,KHgSb_Ma_exp,KHgSb_Liang_exp,Tayrong_KHgX}. Due to diverse chemical bondings, the topological state in the AB$X$ materials is highly sensitive to the materials' parameters and electron interactions~\cite{zunger_TIvsTDS}. The distinction between the topological metal or insulating state is thus not very obvious in these materials.   

KZnBi, which is of particular interest to this study, was predicted theoretically as a $\mathbb{Z}_2$ trivial insulator~\cite{zunger_TIvsTDS} or a metal~\cite{ABC_ferroelectric}. A recent experimental work, however, reported it as a 3D topological Dirac semimetal with a pair of bulk Dirac fermions in the bulk Brillouin zone (BZ)~\cite{KZnBi_prx}. The angle-dependent magneto-transport measurements reveal a phase transition to realize chiral fermion states with anomalous Hall effect under an applied magnetic field~\cite{KZnBi_npj}. The nontrivial features in KZnBi can be tuned by varying the direction and flux of the Berry curvature field~\cite{KZnBi_npj, KZnBi_sciadv}. These results indicate that KZnBi realizes a topological Dirac semimetal state in disagreement with $\mathbb{Z}_2$ trivial insulator or a metallic state reported in the first-principles studies. In this context, the density functional theory (DFT) based first-principles calculations, in principle, can reproduce the ground state of materials accurately provided the exact exchange-correlation (XC) functionals. Proper inclusion of exchange and correlation effects may thus be crucial to describe the topological Dirac semimetal state of KZnBi. 

In this work, we provide a comprehensive investigation of the electronic structure and topological properties of KZnBi based on first-principles calculations. By carefully considering the different XC functionals including the strongly constrained and appropriately normed (SCAN) and hybrid functional with Hartree-Fock (HF) exchange tuning, we delineate the electronic and topological state of KZnBi. Our analysis demonstrates that SCAN density functional generates the structural parameters in agreement with the experimental results. However, it produces a topological nonsymmorphic insulator state in disagreement with the experiments. By tuning the XC strength in the modified Becke-Johnson (mBJ) potential, we recover the correct orbital ordering and Dirac semimetal state of KZnBi. We further show that the Dirac semimetal state can also be captured through an increased Hartree-Fock (HF) exchange in the HSE06 hybrid functional. Our results indicate that the strong exchange and correlation effects are key to realizing the topological Dirac semimetal state of KZnBi in agreement with the experiments.

\section{Methodology}\label{method}
Electronic structure calculations are performed within the Kohn-Sham DFT (KSDFT) framework using the projector augmented-wave (PAW) method to model the ionic potentials as implemented in Vienna {\it ab-initio} simulation package (VASP)~\cite{KSDFT,PAW,VASP1}. The accuracy of the KSDFT depends solely on the XC functionals, which can be arranged along different rungs of Jacob's ladder towards chemical accuracy~\cite{JacobL,DFT_ReviewNM}. The XC functionals are broadly categorized into semilocal, hybrid, and post-hybrid functionals. The semilocal XC functionals can be written as
\begin{equation}
E_{\rm xc}[\rho_\uparrow,\rho_\downarrow]=\int d^3r  \rho({\bf r})\epsilon_{\rm xc}(\rho_\uparrow,\rho_\downarrow, \nabla \rho_\uparrow,\nabla \rho_\downarrow,\tau_\uparrow,\tau_\downarrow)
\end{equation}
where $\rho_{\uparrow}({\bf r})$ and $\rho_{\downarrow}({\bf r})$ are the electron spin densities and constitute the local density approximation (LDA)~\cite{LDA,ground1980ceperley}. Apart from the electron spin densities, spin density gradients ($\nabla\rho_\uparrow({\bf r})$ and $\nabla\rho_\downarrow({\bf r})$) are included in generalized gradient approximations (GGA)~\cite{PBE}. In the meta-GGAs additional kinetic energy densities ($\tau_{\sigma=\uparrow,\downarrow}=\sum_{i=1}^{occ}\frac{1}{2}|\nabla\psi_{i,\sigma}|^2$) are added~\cite{SCAN}. Notably, semilocal XC functionals can explain the equilibrium geometry and symmetries of materials although they can miss the correct location of the highest occupied and lowest unoccupied states due to the inherent self-interaction error~\cite{LDA}. The hybrid functionals are designed to overcome this issue by suitably mixing the semilocal part with the exact HF exchange energy~\cite{rationale1996perdew, hybrid2003heyd}. The range-separated hybrid functional HSE($\alpha, \omega$) can be thus written as
\begin{equation}
E_{\rm xc}^{\rm HSE}=\alpha E_{\rm x}^{\rm SR,HF}(\omega)+(1-\alpha)E_{\rm x}^{\rm SR,PBE}(\omega)+E_{\rm x}^{\rm LR,PBE}+E_{\rm c}^{\rm PBE}
\label{hse}
\end{equation}
where $\alpha$ defines the fraction of exact HF exchange mixed at the short range and $\omega$ determines the range-separation parameter. Another way to correct the band gap at the semilocal level is to directly approximate the exchange potential~\cite{OEP}. In this spirit, modified Becke-Johnson (mBJ) exchange potential is developed which is given as
\begin{equation}
v_{{\rm x},\sigma}^{\rm TB-mBJ}({\bf r})=c v_{{\rm x},\sigma}^{\rm BR}({\bf r})+(3c-2)\frac{1}{\pi}\sqrt{\frac{5}{12}}\sqrt{\frac{2\tau_\sigma({\bf r})}{\rho_\sigma({\bf r})}}
\label{mbj}
\end{equation}
where, $v_{x,\sigma}^{BR}({\bf r})$ is the Becke-Roussel potential~\cite{a2006becke}. The value of parameter $c$ ($c_{mBJ}$, henceforth) can be determined from the average of $\nabla\rho/\rho$ over the unit cell~\cite{TBMBJ}. The first term in Eq.~\ref{mbj} denotes the average HF potential, whereas the second term gives the screening potential. Equation~\ref{mbj} can thus be regarded as a hybrid potential where $c_{mBJ}$ controls the mixing of HF and electron screening potentials.

We used the aforementioned XC functionals lying on different rungs of Jacob's ladder to calculate the electronic and topological state of KZnBi (see Table~\ref{lattice}). The kinetic energy cut-off of 420 eV for the plane-wave basis set and Gaussian smearing with a smearing width of 50 meV were used. We considered the van der Waals corrections within the DFT-D3 method of Grimme with zero damping function~\cite{DFTD3}. The spin-orbit coupling (SOC) was added self-consistently to include the relativistic effects~\cite{PBE}. The optimization of lattice geometry was performed until the forces on each atom were less than $10^{-2}$ eV$\AA^{-1}$. A tolerance for electronic energy minimization as $10^{-6}$ eV and a $\Gamma-$centred $11 \times 11 \times9$ $k$-mesh for the BZ sampling were considered. The topological properties were calculated using the materials-specific tight-binding model generated using the vasp2wannier interface. The atom-centered Wannier functions were generated using the Zn-$s$ and Bi-$p$ orbitals~\cite{mostofi2008wannier90}. We calculated the surface states using the iterative Green's functions method with the WannierTools package~\cite{green1985,WT}.

\section{Results}\label{results}

\subsection{Crystal structure and symmetries}

\begin{figure}[t!]
\includegraphics[width=0.98\linewidth]{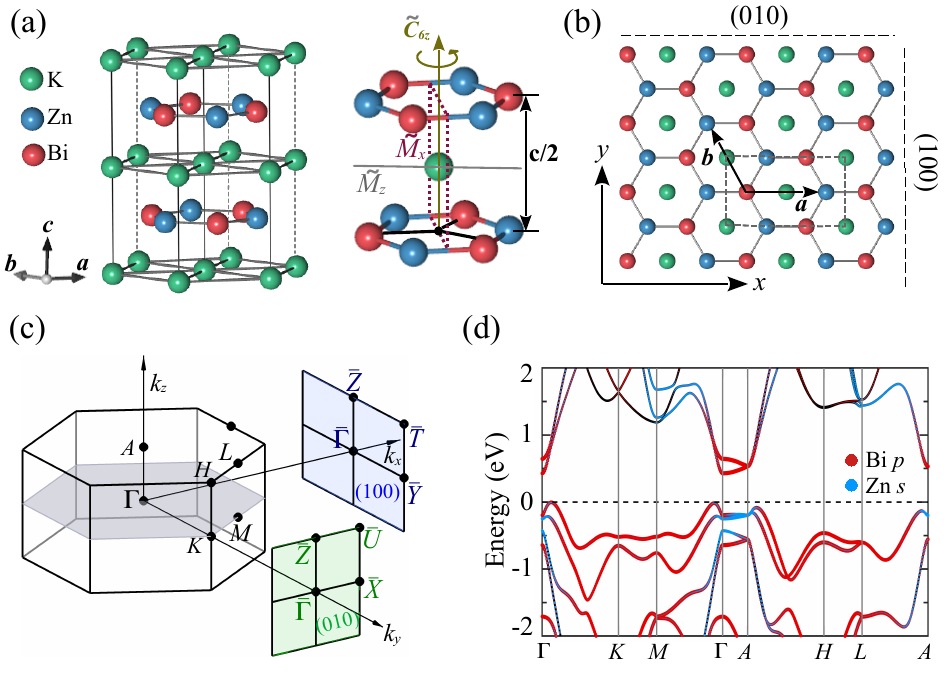}
\caption{{\bf Crystal lattice, symmetries, and band structure of KZnBi.} (a) Bulk crystal structure of KZnBi with the space group P$6_3/mmc$ (left). The ZnBi honeycomb layers mediated by K layers are stacked along the hexagonal $\hat{z}$ axis. The minimal unit cell arrangement and symmetries are shown in the right panel. Maroon dotted plane and solid grey line denote the $\tilde{M}_x$ and $\tilde{M}_z$ mirror planes. The dark yellow line shows the three-fold rotational axis. (b) Top view of the crystal structure with truncated zigzag and armchair directions. (c) Bulk Brillouin zone (BZ) and associated (100) (blue color), and (001) (green color) plane projected surface BZs. The high-symmetry points are marked. The grey plane inside the bulk BZ denotes the $\tilde{M_z}$ mirror plane. (d) Calculated band structure of KZnBi with SCAN+vdW density functional using fully optimized structural parameters. Red and blue colors indicate the contribution from the Bi-$p$ and Zn-$s$ states.}
\label{fig1}
\end{figure}

KZnBi belongs to the non-symmorphic space group P$6_3/mmc$ (No. 194; $D^4_{6h}$) and consists of ZnBi planer honeycomb layers with K atoms lying between them~\cite{KZnBi_prx}. The \{$-A-B-$\}$_n$ stacking of the ZnBi honeycomb layers with weak interlayer coupling mediated by K atoms in the $\hat{z}$ direction forms the KZnBi crystal structure (see Fig.~\ref{fig1}(a)). The K, Zn, and Bi atoms occupy Wyckoff positions $2a~(0, 0, 0)$, $2d~(\frac{1}{3}, \frac{2}{3}, \frac{3}{4})$, and $2c~(\frac{1}{3}, \frac{2}{3}, \frac{1}{4})$, respectively,  in the lattice.  Upon considering the valence electron configuration of the K, Zn, and Bi atoms, the K atoms form a monovalent cationic state (K$^+$) with one electron transferred to the honeycomb lattice ([ZnBi]$^-$). Such a charge transfer satisfies the octet rule of electron filling and suggests that KZnBi may realize a small gap insulator or band overlap semimetal~\cite{APapoian2000}. 

The crystal symmetries of KZnBi include an inversion $\mathcal{I}$ centred at K atom, six-fold screw rotation $\tilde{C}_{6z}$: $(x,y,z)\rightarrow (x \ cos(\theta)-y \ sin(\theta), x \ sin(\theta)+y cos(\theta),z+c/2)$ with $\theta=2\pi/6$, a glide reflection $\tilde{M}_x$: $(x,y,z)\rightarrow (-x,y,z+c/2)$, $M_y$: $(x,y,z)\rightarrow (x,-y,z)$, and $\tilde{M_z}$: $(x,y,z)\rightarrow (x,y,-z+1/2)$. Note that $\tilde{M}_x$ and $\tilde{C}_{6z}$ are non-symmorphic symmetries that involve unremovable lattice translation even by shifting the origin of the unit cell. The top view of the crystal structure is shown in Fig.~\ref{fig1}(b) with a transformed orthorhombic unit cell. The (100) and (010) planes (surface normal is given in the cartesian coordinate system) represent the truncated ZnBi honeycomb lattice in zigzag and armchair directions. The bulk BZ and its associated (010) and (100) surface BZs of KZnBi are shown in Fig.~\ref{fig1}(c). The grey plane centered at $k_z=0$ shows the $\tilde{M_z}$ mirror plane. A detailed discussion of surface symmetries and their associated electronic features is given below.

We present the optimized lattice parameters of KZnBi obtained with different XC functionals in Table~\ref{lattice}. The parameters obtained with the LDA are smaller than the corresponding experimental values. The calculated bond lengths of in-plane Zn-Bi bonds ($d_{Zn-Bi}$) and out-of-plane K-\{ZnBi\} bond ($d_{K-ZnBi}$) are smaller than their associated experimental values. This is expected since the LDA is known to overestimate bond strengths in materials. The lattice parameters obtained with PBE-GGA are overestimated almost by the same amount as the LDA underestimated them. Since KZnBi has a layered structure, the van der Waals (vdW) interaction plays a significant role in its structure parameters. The mismatch in lattice parameters with experimental values decreases when vdW interactions are included during geometry optimization with PBE-GGA~\cite{DFTD3}. In contrast, SCAN, and R2SCAN meta-GGA functionals provide lattice parameters close to their experimental values. Importantly, the SCAN+vdW generates parameters having an excellent match with the corresponding experimental values. We thus describe the electronic properties of KZnBi using the SCAN+vdW optimized parameters in our later discussion. 

\begin{table}[t!]
\caption{Calculated lattice parameters for KZnBi in P$6_3/mmc$ (No. 194; $D^4_{6h}$) space group using different XC functionals.  $a$ and $c$ are the hexagonal in-plane and out-of-plane lattice constants. $d_{Zn-Bi}$ and $d_{K-ZnBi}$ are bond lengths of in-plane Zn-Bi and out-of-plane K-\{ZnBi\} bonds. }
\begin{tabular}{l c c c c}
\hline\hline
				  &  $a$ (\AA)  	&~~   $c$ (\AA)  	& $d_{Zn-Bi}$ (\AA)  &  $d_{K-ZnBi}$ (\AA)  \\
\hline
LDA                           	&  4.591  		&~~  10.262    	             & 	2.650    	&  	2.565 	\\
PBE-GGA              	&  4.743  		&~~ 	10.913  	& 	2.738   		&  	2.728  	\\
PBE+vdW                    &  4.711 		&~~ 	10.692   	& 	2.720   		& 	2.673  	\\
SCAN                           &  4.652 		&~~ 	10.706 	            & 	2.685  		& 	2.676   	\\
R2SCAN                      &  4.678 		&~~ 	10.725  	& 	2.700   		&  	2.681		\\
SCAN+vdW                 &  4.636  		&~~ 	10.601  	& 	2.676    	& 	2.650		\\
\hline 
Exp.~\cite{KZnBi_prx}   & 4.676        &~~	10.597	& 	2.699  		 &	2.649 	\\
\hline\hline
\end{tabular}
\label{lattice}
\end{table}

\subsection{Topological non-symmorphic crystalline insulator state}

\begin{figure}[t!]
\includegraphics[width=0.98\linewidth]{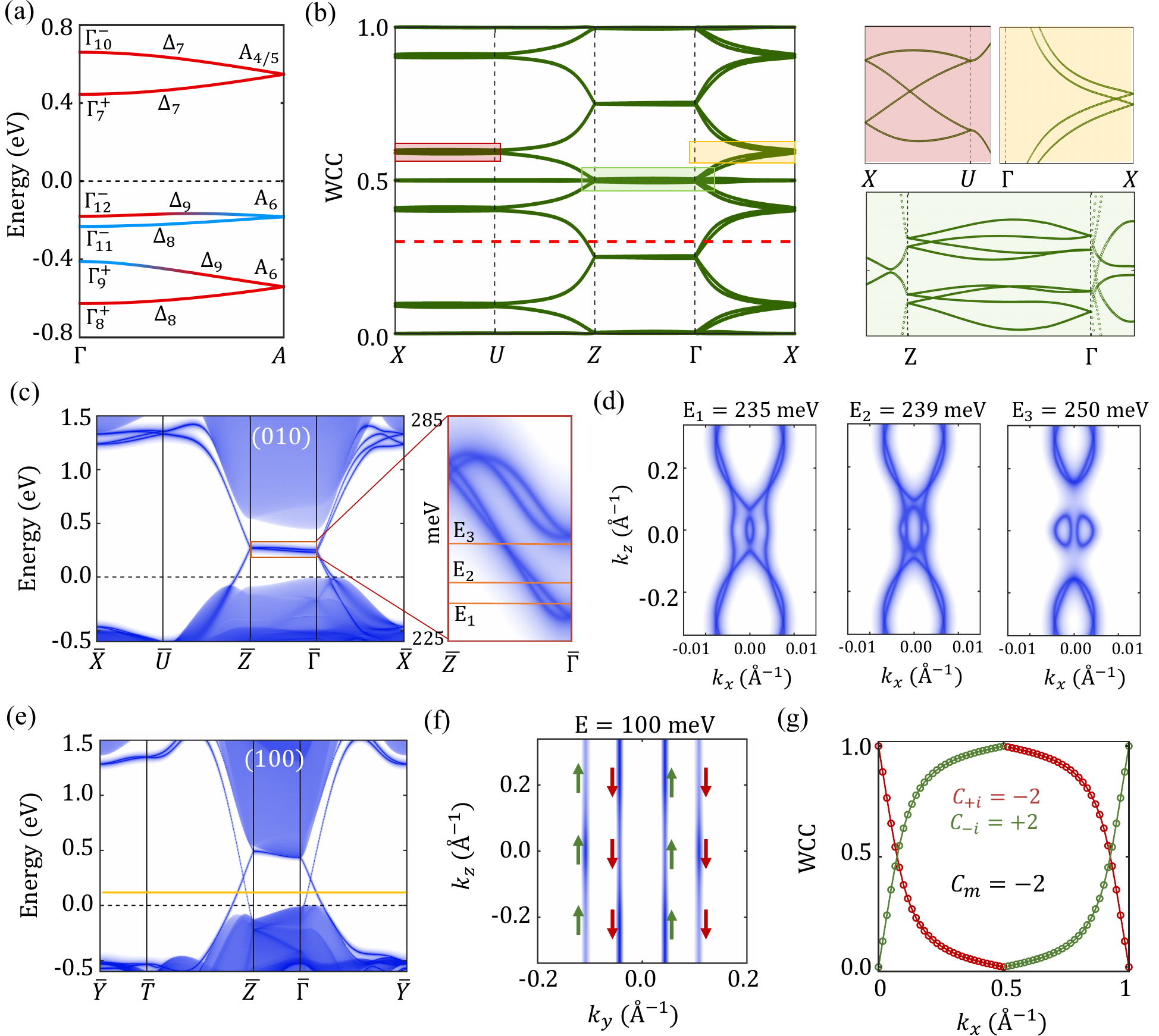}
\caption{{\bf Topological non-symmorphic crystalline insulator state of KZnBi.} (a) Closeup of the bulk bands along the $\Gamma-A$ line with irreducible representations. The Bi-$p$ and Zn-$s$ orbital characters are given in red and blue, respectively. (b) Calculated Wannier charge centers (WCC) spectrum along various high-symmetry lines. The dashed red line identifies an arbitrary reference line to calculate the topological invariants (see text for details). Closeups of the spectrum in the light red, yellow, and green boxes are given in the insets. (c) Band structure of the (010) surface of KZnBi.  The inset shows a zoomed-in view of the surface Hourglass fermion. (d) Constant surface band contours at various energies. (e)  Band structure of the (100) surface of KZnBi. (f) Surface band contour at 100 meV (yellow line in (e)) with the associated spin texture. The four counter-propagating spin-polarized states represent the double quantum spin-Hall states. (g) Evaluation of the WCCs on the $k_z=0$ plane with nonzero mirror Chern number $C_m=-2$.}
\label{fig2}
\end{figure}

The bulk band structure of KZnBi using fully relaxed structural parameters is shown in Fig.~\ref{fig1}(d). It is insulating with an indirect band gap of 445 meV. The bands near the Fermi level consist of strongly hybridized Bi-$p$ and Zn-$s$ states with a band inversion at the $\Gamma$ and $A$ points. The presence of both time-reversal and inversion symmetries in KZnBi enforces Kramers's degeneracy in all the bands. The non-symmorphic symmetries enforce additional degeneracy in the valence and conduction bands at the BZ boundaries. Such band degeneracies are evident along the $L-A$ symmetry line in Fig.~\ref{fig1}(d). To discuss the topological state, we present the irreducible representations of the space group characterizing band symmetries along the $\Gamma-A$ line in Fig.~\ref{fig2}(a). The bands at $\Gamma$ and $A$ points are two and four-fold degenerate, respectively. At $A$ point, there are two types of representations, $A_6$ and $A_4+A_5$ ($A_{4/5}$, in short). As one moves away from A to $\Gamma$ at any intermediate point $\Delta$ (0, 0, $\delta k_z$), the four-fold bands split into two two-fold bands $A_6  \rightarrow \Delta_8 +\Delta_9$ and $A_4+A_5 \rightarrow  \Delta_7 +\Delta_7$. At the  $\Gamma$ point, the three $\Delta$ point representations evolve to $\Gamma_i^{\pm}$ ($i =7-12$), where $\pm$ represents parity eigenvalue of the states. Importantly, $A_6$ and $\Gamma_{11,12}^{-}$ bands are inverted and lie in the valence region without having any band crossings along the $\Gamma-A$ direction. The parity eigenvalue analysis gives a $\mathbb{Z}_2=0$, revealing band inverted $\mathbb{Z}_2$ trivial insulator state of KZnBi. 

To characterize the exact topological state of KZnBi, we present the Wannier charge centers (WCC) evolution of the occupied states along $X-U-Z-\Gamma-X$ direction in Fig.~\ref{fig2}(b). The nontrivial connection in the WCC spectrum is clearly seen. Note that the glide reflection $\tilde{M}_x$ combined with the time-reversal $\mathcal{T}$ symmetry defines a $\mathbb{Z}_4$ invariant ($\chi$) while the mirror operator $\tilde{M}_z$ ensures a well-defined mirror Chern number (MCN) $C_m$ on the $k_z=0$ and $\pi/c$ planes. The relation between the WCC spectrum and topological invariants is given as~\cite{wallpaper_fermion,Tayrong_KHgX},
 
 \begin{equation}
\chi=2(n_{XU}^+ + n_{Z\Gamma}^+) +n_{\Gamma X}~{mod}~4
\label{z4}
\end{equation}
and
\begin{equation}
C_m= (n_{-X\Gamma X}^{+i} - n_{-X\Gamma X}^{-i})/2
\label{mcn}
\end{equation}

The numbers $n$'s in Eq.~\ref{z4} and \ref{mcn} can be evaluated from the WCC spectrum as follows.
If we choose the $\tilde{M}_x$ glide subspace with eigenvalue $+i e^{(-i k_z c/2)}$ along $X \rightarrow U$ and $Z \rightarrow \Gamma$ direction, $n_{XU}^+$ and $n_{Z\Gamma}^+$ is defined as the difference between the number of times an arbitrary horizontal reference line crosses the bands with positive and negative slopes.  $n_{\Gamma X}$ is the difference between the number of times the reference line crosses positive and negative slope bands in the $\Gamma \rightarrow X$ direction. Similarly, in the $\tilde{M_z}$ mirror subspace along $-X \rightarrow \Gamma \rightarrow X$ direction, $n_{-X\Gamma X}^{+i}$ and $n_{-X\Gamma X}^{-i}$ gives the difference between the number of times this line crosses the positive and negative slope bands. Since the system is time-reversal symmetric, $n_{-X\Gamma X}^{+i} = - n_{-X\Gamma X}^{-i}$. Note that while taking this difference, one should compute \{positive slope\} - \{negative slope\} bands. Following this convention, the values deduced from the WCC spectrum are $n_{XU}^+=0$,  $n_{Z\Gamma}^+=0$, $n_{\Gamma X}=2$ and $n_{-X\Gamma X}^{+i} =-2$.  The obtained topological invariants are $\chi=2$ and $C_m=-2$. This shows that KZnBi is a topological non-symmorphic crystalline insulator similar to KHgX~\cite{Hourglass_bernevig,KHgSb_Ma_exp,KHgSb_Liang_exp,Tayrong_KHgX}. The topological state of KZnBi is further evaluated by calculating the electronic structure and WCCs spectrum using different XC functionals (LDA, PBE-GGA, SCAN, R2SCAN) including the HSE06 functional with 25\% exact HF exchange (results not shown for brevity). All these functionals produce a  topological non-symmorphic crystalline insulator. 
   
The nontrivial surface states in topological crystalline insulators (non-symmorphic and symmorphic) appear only on the crystal surfaces that respect crystalline symmetries protecting the topological state. To showcase these nontrivial states, we calculate the surface band structure of (010) and (100) surfaces of KZnBi [Figs.~\ref{fig2}(c) and \ref{fig2}(e)]. On the (010) surface, the glide-mirror $\tilde{M}_x$ and mirror $\tilde{M}_z$ are preserved. Note that the time-reversal $\mathcal{T}$ and $\tilde{M}_x$ symmetries commute $i.e.$ $[\mathcal{T}, \tilde{M}_x]=0$. This results ($\mathcal{T} \tilde{M}_x)^2$ =$\mathcal{T}^2$ $\tilde{M}_x^2$ = t(c$\hat{z}$)=$e^{-ik_z c}$. At $k_z=\pi/c$, $(\mathcal{T} \tilde{M}_x)^2=-1$ and thus, a Kramer-like double degeneracy is enforced at every wave vector ($\textbf{k}$) on the  $\overline{Z}-\overline{U}$ line as seen in Figs.~\ref{fig2}(c). We observe Mobius twisted bands along the $\tilde{M}_x$ invariant $\overline{X}-\overline{U}$ and $\overline{Z}-\overline{\Gamma}$ paths. The eigenvalue of $\tilde{M}_x$ operator at any wave vector along these lines is $\pm i e^{-ik_z c/2}$. When $k_z=0$, the eigenvalues are $\pm i$ and when $k_z=\pi/c$, the eigenvalues are $\pm 1$. This eigenvalues switching enforces bands to come in pair of four, giving rise to hourglass surface fermions in KZnBi. The hourglass band dispersion is clearly revealed in the zoom-in view of bands along $\overline{Z} - \overline{\Gamma}$ in Fig.~\ref{fig2}(c). The evolution of hourglass surface states in the 2D surface BZ is shown in Fig.~\ref{fig2}(d). 

To resolve the $\tilde{M_z}$ mirror symmetry protected surface states, we present the calculated band structure of the (100) surface in Fig.~\ref{fig2}(e). On the $k_z=0$ plane (gray plane in Fig.~\ref{fig1}(c)), the bulk bands split into $+i$ or $-i$ subspace of $\tilde{M_z}$. By decomposing the Hamiltonian $H(k)= H^{+ i}(k)+H^{-i}(k)$, where $H^{\pm i}(k)$ is spanned by the mirror eigenstates $\ket{\psi^{\pm i}_{k,n}}$ satisfying $\tilde{M_z} \ket{\psi^{\pm i}_{k,n}}= \pm{i} \ket{\psi^{\pm i}_{k,n}}$, and calculating the associated Chern numbers, we obtain the mirror Chern number $C_{m}=-2$ (Fig.~\ref{fig2}(g)). Due to this nontrivial $C_m$ on the $k_z=0$ plane, topological nontrivial Dirac cone states are seen. However, these bands are almost non-dispersive along the $k_z$ direction (${\overline{\Gamma}}-\overline{Z}$ line), resulting in a highly anisotropic band dispersion with a nearly flat band dispersion along ${\overline{\Gamma}}-\overline{Z}$. We show the constant energy contours at 100 meV above the Fermi level on the $k_y-k_z$ plane in Fig.~\ref{fig2}(f). The two pairs of spin-polarized counter-propagating surface states that are well-spaced in momentum space are clear. These states reflect the double quantum spin-Hall insulator state in KZnBi~\cite{Hourglass_bernevig}. 

\subsection{Electronic state tuning and topological Dirac semimetal state}

\begin{figure}[h!]
\includegraphics[width=0.98\linewidth]{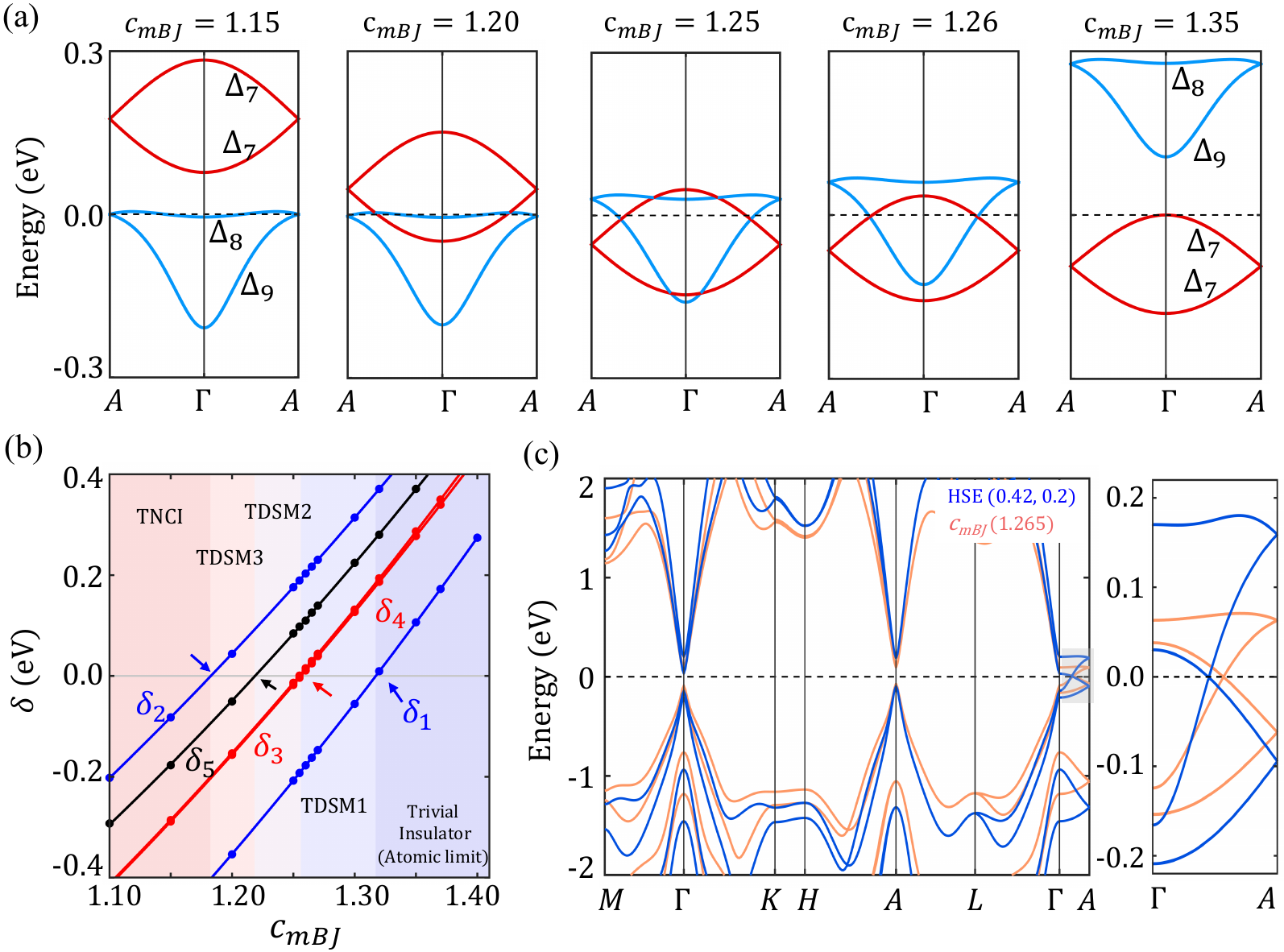}
\caption{{\bf Evolution of band crossings and topological phase tuning in KZnBi.} (a) Calculated band structure along $A-\Gamma-A$ line for various values of $c_{mBJ}$. The space group representations are noted on bands. (b) Band gap at $\Gamma$ ($\delta_{i,~{i=1-4}}$) and $A$ points as a function of $c_{mBJ}$ (see text for details). The horizontal grey line at $\delta =0$ marks the reference line separating the normal and inverted states. The solid markers represent the calculated values, whereas solid lines are guides to the eye. TDSM1, TDSM2, and TDSM3 represent topological Dirac semimetal with different ordering of the crossings bands. TNCI represents a topological non-symmorphic crystalline insulator. (c) Bands structure obtained with $c_{mBJ}=1.265$ and HSE(0.42, 0.2) hybrid functional with $42\%$ exact HF exchange (see Eq.~\ref{hse}). Closeup of the bulk bands along $\Gamma-A$ is shown in the right panel.}
\label{fig3}
\end{figure}

The preceding analysis demonstrates that KZnBi realizes a topological nonsymmorphic crystalline insulator with different XC functionals in their default setting. This is in disagreement with the recent ARPES results where a topological Dirac semimetal state is seen~\cite{KZnBi_prx}. To resolve this issue, we now present the bulk bands obtained by changing the parameter $c_{mBJ}$ of mBJ potential in Fig.~\ref{fig3}. This parameter mixes the average HF and screening potential and can be tuned manually to obtain the correct band ordering~\cite{TBMBJ, mbJ_limitation}. Figure~\ref{fig3}(a) presents the evolution of bulk bands with $c_{mBJ}$ along $A-\Gamma-A$ line. In the atomic limit of KZnBi, the space group representations of the valence bands are $\Delta_7$ and $\Delta_7$, whereas the conduction band is $\Delta_8$ and $\Delta_9$ at an intermediate point $\Delta$ on this line. The associated $\tilde{C}_{6z}$ rotational eigenvalues are

\begin{equation}
 \begin{aligned}
 \Delta_7=\{e^{i\pi/2},e^{-i\pi/2} \}\\
 \Delta_8=\{e^{i5\pi/6},e^{-i5\pi/6} \} \\
 \Delta_9=\{e^{i\pi/6},e^{-i\pi/6} \} 
  \end{aligned}
\end{equation}

Here, we omit the phase factor $e^{(-ik_z c/2)}$ arising from the lattice translation along $c$ direction. Due to inversion and time-reversal symmetries, these eigenvalues are degenerate in pairs. 
The band crossings between different $\Delta_i$ bands will not hybridize and can result in symmetry-protected band crossings. At the $c_{mBJ}$ value of 1.15, $\Delta_8$ and $\Delta_9$ bands lie in the valence region, developing double band inversions at both $\Gamma$ and $A$ points. With an increase in $c_{mBJ}$ to 1.20, $\Delta_7$ and $\Delta_8$ bands cross at the Fermi level to form stable band crossings. With a further increase in $c_{mBJ}$ to 1.26, $\Delta_7$ band crosses with $\Delta_9$ band to generate the stable band crossings at the Fermi level. When $c_{mBJ}$ is increased to 1.35, both $\Delta_7$ bands move to the valence region and generate a trivial insulator state.   

To clarify the distinct ordering of the crossing bands, we define energy gap $\delta_i~(i=1-5)$ at $\Gamma$ and $A$ points as 

\begin{equation}
 \begin{aligned}
 \delta_1= \Gamma_{9}^+ -\Gamma_{10}^- \\
 \delta_2=  \Gamma_{11}^- -\Gamma_{7}^+ \\
 \delta_3=  \Gamma_{9}^+ -\Gamma_{7}^+ \\
 \delta_4=  \Gamma_{11}^- -\Gamma_{10}^- \\
 \delta_5= A_6-A_{4/5} 
  \end{aligned}
\end{equation}

The evolution of these $\delta_{i}$'s as function of $c_{mBJ}$ is shown in Fig.~\ref{fig3}(b). All $\delta_i$'s are negative at lower $c_{mBJ}$ values ($c_{mBJ} < 1.17$), and whenever $\delta_i$'s changes sign, a band inversion gets unlocked. The critical values of $c_{mBJ}$ at the phase transition points are marked by arrows in Fig.~\ref{fig3}(b). At lower $c_{mBJ}$ values, a topological nonsymmorphic crystalline insulator with $\chi =2$ and $C_m=-2$ is formed. At the intermediate values of $c_{mBJ}$, a topological Dirac semimetal state with $C_m=-1$ is realized. However, the ordering of crossing bands that form Dirac states changes with $c_{mBJ}$. At larger values of $c_{mBJ}> 1.32$, a trivial insulator is realized. On comparing with the experimental results, we find that the topological Dirac semimetal phase realized with $1.26<c_{mBJ}<1.32$ better reflects the experimental situation. Such a Dirac semimetal phase and band orderings are further reproduced by increasing the HF exchange from 25\% to 42\% in HSE06 hybrid functional (see Fig.~\ref{fig3}(c)). These results indicate that the electron correlations effects are essential to restore the correct orbital ordering and topological Dirac semimetal state of KZnBi.

\begin{figure}[t!]
\includegraphics[width=0.98\linewidth]{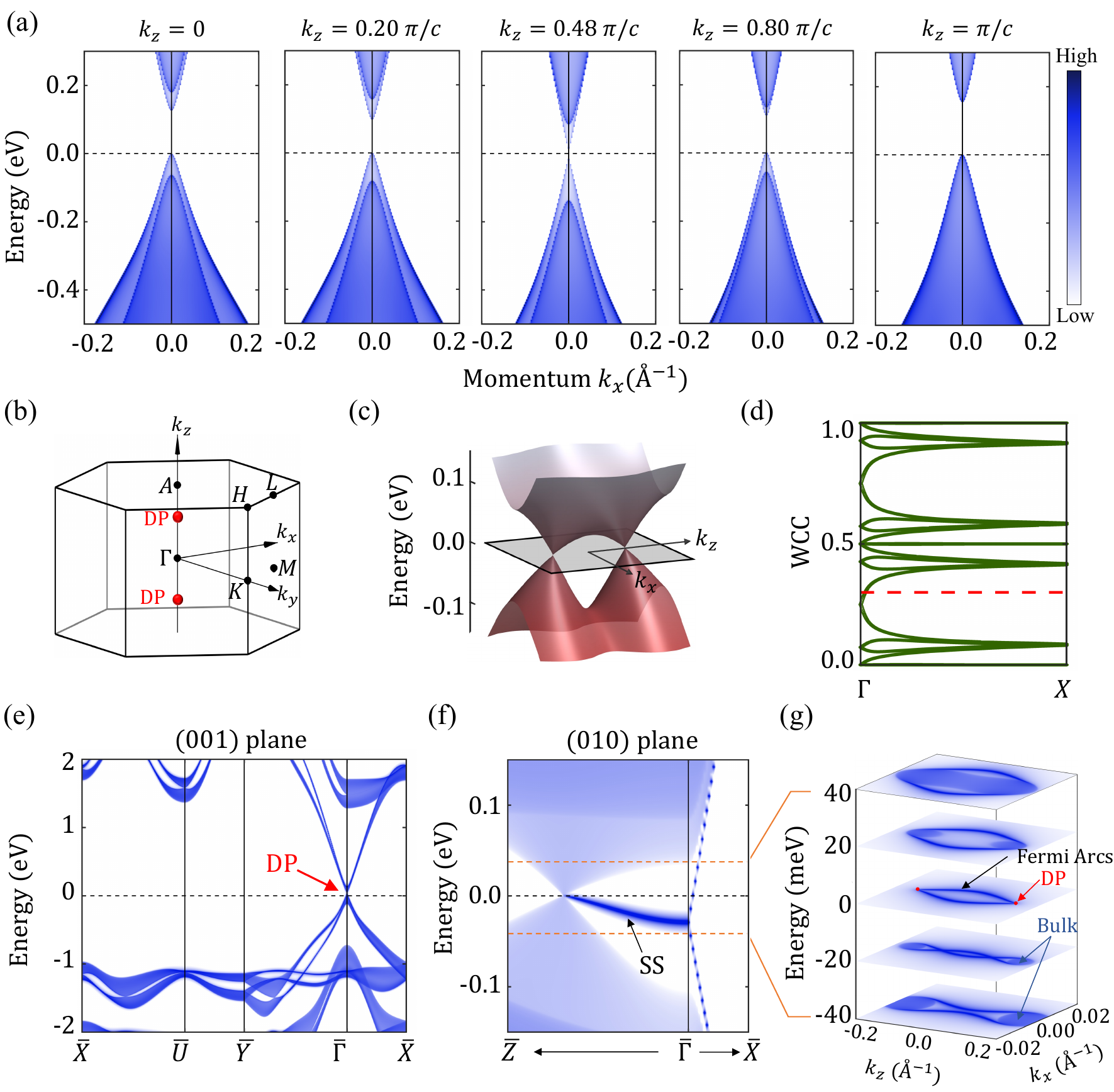}
\caption{{\bf Topological Dirac semimetal state of KZnBi.} (a) Bulk band structure of KZnBi projected on the (010) surface Brillouin zone as a function of $k_z$. (b) Location of the four-fold Dirac nodes in the hexagonal Brillouin zone on the rotational axis (red sphere). (c) Energy dispersion of the Dirac cones in the $k_x-k_z$ plane. (d) Evolution of Wannier charge centers (WCCs) on the $k_z=0$ plane along $\Gamma-X$ line. The WCCs cross the red reference line one time,  revealing nontrivial $\mathbb{Z}_2= 1$. (e)-(f) Calculated band structure of (001) and (010) surface of KZnBi. Topological surface states emanating from the projected Dirac nodes are resolved. (g) Evolution of surface band contours as a function of energy. Double Fermi arc surface states connecting the projected bulk Dirac nodes are resolved at the Fermi energy.}
\label{fig4}
\end{figure}

\begin{figure}[hb!]
\includegraphics[width=0.98\linewidth]{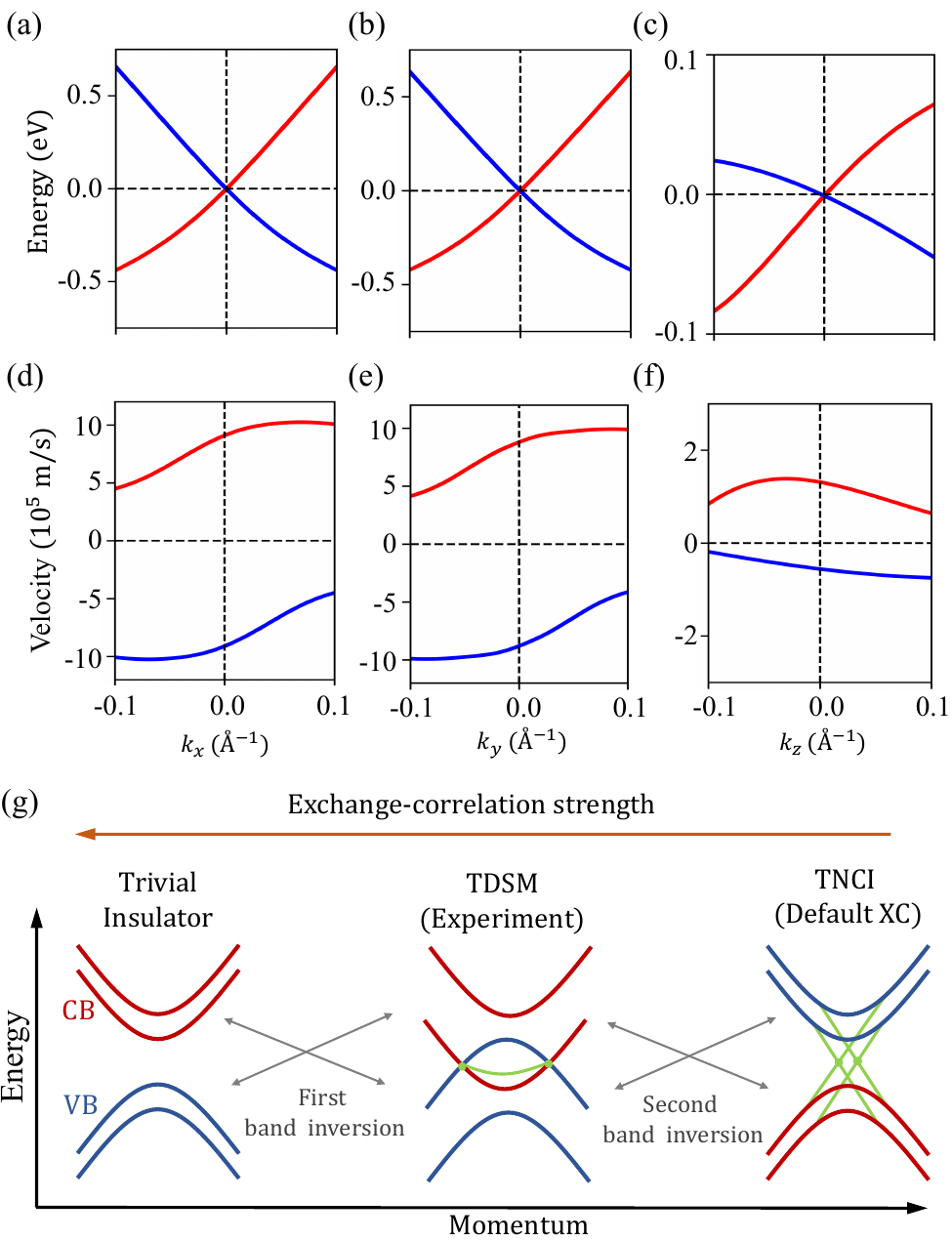}
\caption{ {\bf Energy dispersion and Fermi velocity of Dirac states.} Calculated energy dispersion of the Dirac bands along (a) $k_x$, (b) $k_y$, and (c) $k_z$ directions. The center of the axis is located at a Dirac node. Calculated carrier velocity along (d) $k_x$, (e) $k_y$, and (f) $k_z$ directions. (g) Schematic evolution of topological states on KZnBi as a function of the strength of XC effects. The words “Experiment” and “Default XC” identify the phases observed in experiments (Ref. \cite{KZnBi_prx}) and with default XC strength in various functionals (see text for details).  Blue and maroon curves represent bulk bands and green curves show surface states.}
\label{fig5}
\end{figure}
We discuss the topological properties of the Dirac semimetal state in Fig.~\ref{fig4} and compare them with the experimental results. In Fig.~\ref{fig4}(a), we show the calculated bulk band structure of KZnBi projected on (010) surface BZ for different $k_z$ values. The emergence of a four-fold degenerate Dirac node at $k_z=0.48 \frac{\pi}{c}$ is evident. There are two such symmetry equivalent Dirac nodes on the $A-\Gamma-A$ rotational axis at $(0, 0, \pm 0.48 \frac{\pi}{c})$~\r{A}$^{-1}$ in the bulk BZ (Fig.~\ref{fig4}(b)). These points are clearly resolved in the $E-k_x-k_z$ energy dispersion in Fig.~\ref{fig4}(c), which is highly anisotropic between in-plane and out-of-plane directions. For $k_z \ne 0.48 \frac{\pi}{c}$, a band gap develops between the valence and conduction bands. The presence of bulk Dirac cone at $k_z=\pm 0.48 \frac{\pi}{c}$ and associated energy-momentum dispersion and its evolution with $k_z$ are in remarkable agreement with the experiments ~\cite{KZnBi_prx}. Since energy dispersion on the $k_z =0$ plane is gapped, we can define the $\mathbb{Z}_2$ number similar to insulators. The calculated WCC spectrum reveals a nontrivial $\mathbb{Z}_2= 1$ on this plane. This demonstrates that the Dirac semimetal state of KZnBi is topologically protected, similar to Na$_3$Bi and Cd$_3$As$_2$ Dirac semimetals~\cite{DSM_Na3Bi_SuYang, DSM_Cd3As2}. Figures~\ref{fig4}(e) and \ref{fig4}(f) show the surface band structure of (001) and (010) surfaces of KZnBi. The two Dirac nodes project at the $\overline{\Gamma}$ point on the (001) plane. On the (010) surface, the Dirac nodes project in the $\Gamma-Z$ line. The topological surface states emanating from a projected Dirac node are seen in Fig.~\ref{fig4}(f). The associated constant energy spectrum at various energies in Fig.~\ref{fig4}(g) reveals the double Fermi arcs states that emanate and terminate at the projected Dirac nodes at the Fermi level. The Fermi arc states change topology as one moves away from the Fermi level. 

We present the local energy dispersion of Dirac bands along $k_x$, $k_y$, and $k_z$ directions in Figs.~\ref{fig5}(a)-(c) and associated Fermi velocities $v_f=\partial E/\partial{k} $ in Figs.~\ref{fig5}(d)-(f). The Fermi velocity of bands at the Dirac nodes is $9.09 \times 10^5 \ m/s$, $8.78 \times 10^5 \ m/s$, and $1.31 \times 10^5 \ m/s$ along $k_x$, $k_y$ and $k_z$ directions, respectively. The energy dispersion and Fermi velocity reveal the anisotropic nature of Dirac carriers, which are in substantial accord with their corresponding experimental results~\cite{KZnBi_prx}. These results establish that enhanced XC effects are essential to correctly reproduce the topological state of KZnBi. In Fig.~\ref{fig5}(g), we schematically show the evolution of different topological states of KZnBi as a function of XC strength.  The different density functionals in their default settings produce a TNCI state with double bulk band inversion. Increasing the strength of XC effects unlocks a band inversion and drives the system into a TDSM state in agreement with experiments. A further increase in the strength of XC effects restores the trivial band ordering in KZnBi.

\section{Conclusion}\label{summary}

We have investigated the electronic structure of KZnBi within the framework of density functional theory, with a focus on delineating its exact topological state. Our structural optimization based on various levels of XC density functionals shows that the lattice parameters obtained with SCAN+vdW are in excellent agreement with the experimental parameters. The associated electronic state realizes an insulator state with inverted band ordering in bulk BZ. The band symmetries and WCC spectrum show that the system realizes a  topological nonsymmorphic crystalline insulator with topological invariants $\chi=2$ and $C_m=-2$. We further find that the calculated electronic and topological state with various XC functionals (PBE-GGA, SCAN and R2SCAN meta-GGA, and HSE06) is found to be distinct from the recent experiments that report KZnBi as a TDSM. We address this discrepancy by showing that tuning the strength of XC effects in various functionals is essential to produce the topological Dirac semimetal state with correct orbital ordering. Specifically, a higher value of $c_{mBJ}$ parameter that mixes the HF exchange and screening potentials recovers the desired orbital ordering and Dirac semimetal state. These results are further substantiated by the energy dispersion obtained with an enhanced exact HF exchange in the HSE06 hybrid functional. Our calculated Dirac nodes, their location, and associated carrier velocities are found to be in excellent agreement with available experimental results. Our study thus demonstrates that KZnBi is a unique topological material where XC effects are essential to producing the correct orbital ordering and topological Dirac semimetal state of KZnBi.

\section*{Acknowledgements} 
This work is supported by the Department of Atomic Energy of the Government of India under Project No. 12-R$\&$D-TFR-5.10-0100 and benefited from the computational resources of TIFR Mumbai. 

\bibliography{KZnBi}
\end{document}